\documentclass[letterpaper]{article} 
\usepackage{aaai25}  
\usepackage{times}  
\usepackage{helvet}  
\usepackage{courier}  
\usepackage[hyphens]{url}  
\usepackage{graphicx} 
\usepackage{natbib}  
\usepackage{caption} 
\usepackage{algorithm}
\usepackage[switch]{lineno}
\usepackage{makecell}
\usepackage{multirow}
\usepackage{array}
\usepackage{amsmath}
\usepackage{amsthm}
\usepackage{booktabs}
\usepackage{graphicx}
\usepackage{bm}
\usepackage[noend]{algpseudocode}
\usepackage{amsfonts}
\usepackage{enumitem}
\usepackage[table]{xcolor}
\usepackage{subfigure}
\newcommand{\baby}{MRFF}

\usepackage{newfloat}
\usepackage{listings}
\DeclareCaptionStyle{ruled}{labelfont=normalfont,labelsep=colon,strut=off} 
\floatstyle{ruled}
\newfloat{listing}{tb}{lst}{}
\floatname{listing}{Listing}
\title{Multifaceted User Modeling in Recommendation: \\A Federated Foundation Models Approach}

\author {
Chunxu Zhang\textsuperscript{\rm 1,\rm 2},
Guodong Long\textsuperscript{\rm 3},
Hongkuan Guo\textsuperscript{\rm 4},
Zhaojie Liu\textsuperscript{\rm 4},
Guorui Zhou\textsuperscript{\rm 4},
Zijian Zhang\textsuperscript{\rm 1,\rm 2},
Yang Liu\textsuperscript{\rm 5},
Bo Yang\textsuperscript{\rm 1,\rm 2}\thanks{Corresponding author.}
}
\affiliations {
\textsuperscript{\rm 1} College of Computer Science and Technology, Jilin University, China\\
\textsuperscript{\rm 2} Key Laboratory of Symbolic Computation and Knowledge Engineering of Ministry of Education, China\\
\textsuperscript{\rm 3} Australian Artificial Intelligence Institute, FEIT, University of Technology Sydney\\
\textsuperscript{\rm 4} Kuaishou Technology\\
\textsuperscript{\rm 5} Institute for AI Industry Research, Tsinghua University\\
\{zhangchunxu, zhangzijian, ybo\}@jlu.edu.cn,
guodong.long@uts.edu.au,
\{guohongkuan, zhouguorui\}@kuaishou.com,
liuzj03@gmail.com,
liuy03@air.tsinghua.edu.cn
}

\begin{document}

\maketitle

\begin{abstract}
Multifaceted user modeling aims to uncover fine-grained patterns and learn representations from user data, revealing their diverse interests and characteristics, such as profile, preference, and personality. Recent studies on foundation model-based recommendation have emphasized the Transformer architecture's remarkable ability to capture complex, non-linear user-item interaction relationships. This paper aims to advance foundation model-based recommendersystems by introducing enhancements to multifaceted user modeling capabilities. We propose a novel Transformer layer designed specifically for recommendation, using the self-attention mechanism to capture sequential user-item interaction patterns. Specifically, we design a group gating network to identify user groups, enabling hierarchical discovery across different layers, thereby capturing the multifaceted nature of user interests through multiple Transformer layers. Furthermore, to broaden the data scope and further enhance multifaceted user modeling, we extend the framework to a federated setting, enabling the use of private datasets while ensuring privacy. Experimental validations on benchmark datasets demonstrate the superior performance of our proposed method. Code is available.
\end{abstract}

%
\begin{links}
    \link{Code}{https://github.com/Zhangcx19/AAAI-25-MRFF}
\end{links}

\section{Introduction}
Foundation model-based recommendation methods~\cite{harte2023leveraging,liu2023pre,wu2023survey,zheng2024harnessing,wu2024survey} have emerged as a promising paradigm in recommender systems. The central idea is to leverage powerful foundation models, pre-trained on extensive datasets, and adapt them to specific recommendation tasks. This is typically accomplished by fine-tuning the models or incorporating them as key components within the recommendation architecture. By harnessing the rich knowledge encoded in the foundation models, these systems are able to deliver more accurate and diverse recommendations. However, the deployment of foundation model-based recommender systems necessitates the aggregation of large-scale user-item interaction data, which introduces significant privacy concerns, including the risk of sensitive information leakage. Addressing these privacy challenges is crucial for the deployment of such systems in real-world settings.

Federated Foundation Models (FFMs)~\cite{zhuang2023foundation,yu2023federated} offer an innovative approach to addressing the privacy and security challenges associated with traditional foundation model development. The key idea behind FFMs is to enable the training of foundation models on distributed data sources while avoiding the sharing of raw user data. This collaborative learning approach has emerged as a pioneering technique within the foundation model domain, providing a promising path for the development of powerful yet privacy-preserving AI systems. However, two critical challenges arise when constructing FFMs for recommender systems: \textbf{First}, existing FFMs generally optimize from a large-scale pre-trained model, which presents significant deployment and computation burdens for portable devices of recommendation users. \textbf{Second}, the naive FFMs regard clients equally and learn the same model for all users, which contradicts the goal of delivering customized recommendations for end users.

This paper introduces the \textbf{M}ultifaceted user modeling in \textbf{R}ecommendations with \textbf{F}ederated \textbf{F}oundation models (\textbf{\baby}), to probe the new FFMs paradigm for recommendations. Specifically, we implement a lightweight foundation model on each client, training it from scratch to reduce the high deployment and computational overhead from large-scale pre-trained models. Furthermore, we devise a multifaceted user modeling mechanism to jointly learn user-specific and group-level personalization. On the one hand, we privately separate certain model parameters to preserve user preferences. In addition, we design a novel Transformer layer and configure a group gating network on clients to hierarchically partition users into specific groups, enabling the server to perform group-level parameters aggregation to capture user correlations. By incorporating user-specific modeling  and group-level personalization, \baby \space can harness a synergistic effect, improving predictive performance while safeguarding user privacy.

We evaluate the performance of our method on real-world industrial datasets. Experimental results show that our proposed \baby \space can empower various foundation model architectures and achieve superior performance compared to state-of-the-art baselines on widely-used click-through rate (CTR) prediction task. Additionally, we conduct comprehensive experiments to analyze our model's efficacy in terms of user-grouping capability. Furthermore, we demonstrate the practical feasibility of \baby \space by assessing its efficiency and privacy-preserving capabilities. Our results indicate that our method achieves higher optimization efficiency compared to baselines, while maintaining a good trade-off between performance and privacy preservation. In summary, the \textbf{key contributions} of this work are as follows,

\begin{itemize}
    \item We present an efficient federated foundation model for the recommender system, named \baby. It trains a lightweight foundation model on each client from scratch, which reduces the heavy computational burden imposed by large-scale pre-trained models.
    \item We develop a multifaceted user modeling mechanism that enables collaborative learning of both user-specific and group-level personalization, thus fostering the development of prospective recommender systems in a privacy-preserving manner.
    \item Extensive experiments on benchmark datasets reveal the outstanding performance of \baby \space against baselines. Furthermore, in-depth analysis further validate its strong practical feasibility.
\end{itemize}

\section{Related Work}
\subsection{Multifaceted User Modeling}
Multifaceted user modeling~\cite{hannon2012multi,mcauley2013hidden} aims to capture complex user preferences by considering a wide range of user data and dimensions. This approach goes beyond traditional user modeling, which often relies solely on historical interactions. By integrating diverse data sources, including social connections, contextual information, and content preferences, multifaceted user modeling creates a more nuanced and dynamic representation of users. This approach has become crucial in recommendersystems~\cite{du2021modeling,li2021leveraging,liu2023user,zheng2024mirror,zheng2024bilateral} to enhance the accuracy and personalization of recommendations. The core principle is that users are not static entities with singular preferences. Instead, they possess multiple facets, and their preferences evolve across time and context. By integrating information about users' behaviors, interests, and interactions across different scenarios, multifaceted user modeling enables more accurate prediction of user needs and delivers more personalized and serendipitous recommendations.

\subsection{Foundation Models for Recommendations}
Foundation Models for Recommendations (FM4RecSys) have emerged as a transformative approach in recommender systems, leveraging the vast knowledge and intricate architectures of Foundation Models (FMs) to enhance personalized content delivery and user experience~\cite{lin2023can,huang2024foundation,zhao2024recommender,liu2024largelanguagemodelenhanced}. The core motivation behind integrating FMs into recommender systems stems from their ability to decipher complex patterns and generalize effectively to unseen data. Research in FM4RecSys encompasses a variety of models, including language foundation models~\cite{zhang2023recommendation,bao2023bi}, multi-modal foundation models~\cite{geng2023vip5,zhou2023exploring}, and personalized agents powered by FMs~\cite{zhang2024generative,zhang2024agentcf}. They focus on pre-trained and fine-tuned approaches, prompting techniques, and leveraging the robust representation and generalization capabilities of FMs for recommendations. FM4RecSys is a significant advancement in the field, offering new opportunities for improving recommender systems' accuracy, interactivity, and ability to provide insightful explanations. As research in this area continues to evolve, it is expected to further reshape the landscape of personalized recommendation services.

\subsection{Federated Foundation Models}
Federated Foundation Models (FFMs)~\cite{ijcai2023p393,yu2024federated,ijcai2024p980} introduce an innovative approach that combines the power of FMs with the privacy-preserving framework of Federated Learning (FL)~\cite{mcmahan2017communication,yan2023personalization,xu2024pefad,miao2025icde}. This integration stems from advancements in AI, particularly through models like LLaMA, BERT, and GPT, which leverage vast amounts of data for pre-training~\cite{yu2024federated,ren2024advances,charles2024towards}. However, optimizing these models typically requires access to sensitive data, which raises significant privacy concerns. FFMs tackle these challenges by utilizing the collaborative learning capabilities of FL to enhance FM performance across multiple end-users while preserving data privacy. This approach enables the development of more personalized and context-aware models, effectively addressing challenges related to data scarcity, computational resources, privacy concerns, and ethical considerations. FFMs provide a flexible and scalable framework for training models while preserving privacy, laying the groundwork for further advancements in both FM training and federated learning. Despite these advantages, current FFMs often depend on large-scale pre-trained models for optimization, presenting significant deployment and computational challenges for resource-constrained devices typically used in recommendersystems. To address this limitation, this paper introduces a novel FFM designed specifically for recommendations. Our method reduces computational overhead by training a lightweight foundation model on each user device from scratch and incorporates both user-specific and group-level personalization modeling, enabling effective modeling of user preferences while preserving privacy.

\section{Preliminary}
Recommendation tasks often exhibit sequential dependencies, where users' current interactions are influenced by their past behaviors. \textbf{Transformer} architectures, with their self-attention mechanisms, are particularly well-suited to capture and model these temporal dependencies effectively. Given user $u$'s interaction history $\{i_1, i_2, ..., i_T\}$, the goal is to recommend the next item at time $T+1$. We decompose the transformer-based recommendation architecture into three distinct components: the \textbf{\textit{Input Layer}}, the \textbf{\textit{Transformer Block}}, and the \textbf{\textit{Prediction Layer}}. Each component will be examined in detail below.

\textbf{Input Layer.} 
Sequential interactions are converted into embedding vectors using item embedding tables. Each item $i$ is represented by either indicator attributes (e.g., item ID) or descriptive characteristics (e.g., item type). The item embeddings for each item in the sequence are gathered using embedding tables, forming the input embedding as follows,
\begin{equation}
    i^{emb} = {\rm Concat}({E_i^m}_{m=1}^{|A_i|})
\end{equation}
where $E_i^m$ represents the embedding vector of item $i$ for the $m$-th attribute, $|A_i|$ denotes the total number of attributes, and $Concat$ refers to the concatenation operation. Additionally, sequence embedding vectors typically incorporate learnable positional embeddings~\cite{kang2018self} or relative positional information~\cite{zhai2024actions} to encode positional relationships.

\textbf{Transformer Block.}
The transformer block consists of two modules: the self-attention module (Attention) and the feed-forward network (FFN). Given input $x_{in}$, the transformer block first passes it through the Attention layer to adaptively aggregate embeddings from all previous items. Next, an FFN is applied to introduce non-linearity and capture interdependencies across latent dimensions. This process is formulated as follows, 
\begin{equation}
    x_{out} = {\rm FFN}({\rm Attention}(x_{in}))
\end{equation} 
To mitigate overfitting and training instability, common techniques such as residual connections~\cite{he2016deep}, normalization~\cite{ba2016layer}, and dropout~\cite{srivastava2014dropout} are typically incorporated into the transformer block.

\textbf{Prediction Layer.}
For each user, we combine the user embedding, candidate item embedding, and latent sequence representations from the transformer blocks as input to a multi-layer perceptron (MLP) for prediction, 
\begin{equation}
    \hat{Y} = {\rm MLP}({\rm Concat}(x_{out}^L, u^{emb}, c^{emb}))
\end{equation} 
where $L$ is the number of transformer blocks. $u^{emb}$ and $c^{emb}$ represent user and candidate item embeddings, respectively, constructed similarly to input sequence embeddings.

\begin{figure*}[!t]
\setlength{\abovecaptionskip}{1mm}
\setlength{\belowcaptionskip}{-3mm}
    \centering
    \includegraphics[width=1.\linewidth]{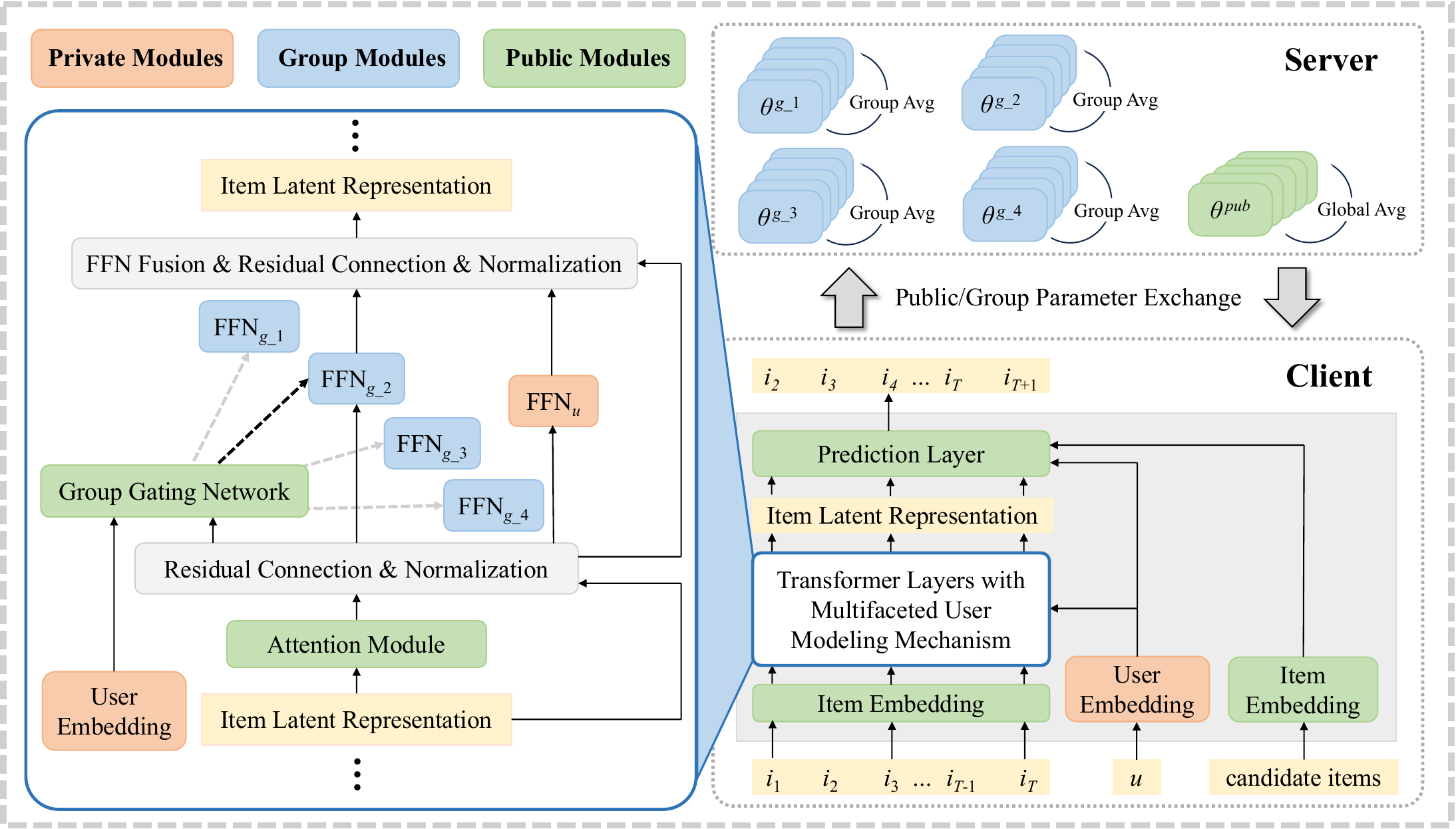}
    \caption{The framework of \baby. The right side illustrates the workflow of our method. Each client trains a lightweight foundation model using personal data. In the transformer layers of the local model, we propose a multifaceted user modeling mechanism to enhance user personalization, with details summarized on the left side. Specifically, we introduce a group gating network after the attention module to direct users to specific FFNs, alongside the user-specific FFN, for forward propagation. During iterative optimization between the server and clients, clients maintain user embeddings and user-specific FFN as private modules to learn user-level personalization. For other parameters, the server aggregates them either globally or by group.}
    \label{framework}
\end{figure*}
\section{Methodology}
\subsection{Overall Framework}
We propose a novel approach for \textbf{M}ultifaceted user modeling in \textbf{R}ecommendations with \textbf{F}ederated \textbf{F}oundation models, referred to as \textbf{\baby}. As shown in Figure~\ref{framework}, \baby \space utilizes a transformer-based foundation model trained on each client, which incorporates our proposed multifaceted user modeling mechanism. To improve personalization, we introduce a group gating network within each transformer layer. This network dynamically directs clients to specific FFNs based on their characteristics. The group-level FFN, together with the user-specific FFN, then contributes to the subsequent feedforward pass. By integrating both user-specific and group-level personalization, \baby \space achieves more effective user modeling while ensuring privacy preservation.

\subsection{Multifaceted User Modeling Mechanism}
Traditional federated learning aims to learn a single shared model across all clients~\cite{mcmahan2017communication}, which is insufficient for recommendation tasks due to the heterogeneity of user data distributions. While users exhibit diverse behaviors, they also share meaningful similarities that can improve individual modeling. Therefore, federated recommender systems require collaboration between learning user-specific parameters and leveraging inter-user relationships. This dual focus on personalization and social connections overcomes the limitations of a one-size-fits-all approach, yielding more robust and effective models.

To achieve this, we propose simultaneously learning personal and group parameters in the federated recommender system. We designate the FFN within each transformer block as a private module to maintain user personalization. This enables the model to capture non-linear relationships in user behavior, addressing the inherent heterogeneity in user data. We also introduce group-level FFNs to model shared patterns across similar users, facilitating federated optimization by leveraging similarities within each user group.

To assign users to their respective FFN modules, we integrate a group gating network after the attention module in each transformer block. The group gating network takes as input the output of the attention module and the user embedding. By combining contextual features learned by the attention mechanism and user-specific information, the group gating network predicts the user's group assignment. We implement the group gating network with an MLP followed by a softmax operation, formalizing the probability of user $u$ belonging to each group as follows,
\begin{equation}
    P_u = softmax({\rm MLP}({\rm Concat}({\rm Attention}_{out}, u^{emb}))
\end{equation}
where ${\rm Attention}_{out}$ represents the output of the attention module in the transformer block. The user's group assignment is determined by selecting the group with the highest probability. This group gating network dynamically assigns users to the appropriate FFNs, effectively matching them with personalization components tailored to their specific behaviors and preferences.

Thus, the computation process within each Transformer block is updated as follows,
\begin{equation}
    x_{out} = {\rm FFN}_u({\rm Attention}(x_{in})) + {\rm FFN}_g({\rm Attention}(x_{in}))
\end{equation} 
where ${\rm FFN}_u$ and ${\rm FFN}_g$ represent the user-specific and group-level FFN, respectively. The inclusion of group gating networks in each transformer block creates a hierarchical model architecture, facilitating the learning of representations at different granularities. Shallow transformer blocks capture general user personas, while deeper blocks reveal more nuanced behavioral segments. This hierarchical approach dynamically assigns users to the appropriate personalization components, effectively balancing shared characteristics with individual preferences.

\subsection{Optimization Objective}
\baby \space aims to build a personalized federated recommender system that delivers tailored recommendations to each user. Each user $u$ stores their personal data $D_u$ locally and trains the foundation model under the server's coordination. The system optimization objective is formulated as follows,
\begin{equation}\label{total_loss}
    \min_{\{\theta_1,...,\theta_n\}} \sum_{u=1}^n \alpha_u \mathcal{L}_u(\theta_u)
\end{equation}
where $\mathcal{L}_u(\theta_u)$ represents the local foundation model loss parameterized by $\theta_u$, and $\alpha_u$ is the weight of user $u$ in the global optimization process.

The local foundation model loss consists of two components. The first component is the recommendation loss, which captures user data characteristics and models individual preferences. We introduce a balance loss to regularize the group gating network, ensuring an even allocation of users into groups. This regularization prevents extreme user groupings, where a few groups dominate the majority of users. We constrain the group gating network to allocate users according to a uniform distribution, 
\begin{equation}
    \mathcal{L}_{balance} = N \cdot \sum_{l=1}^{L} \sum_{i=1}^{N} f_i^l \cdot p_i^l
\end{equation}
where $L$ is the number of transformer blocks and $N$ is the number of FFNs per block. $f_i^l$ and $p_i^l$ represent the proportion of users belonging to the $i$-th group among the total users and the probability of the user being predicted to belong to the $i$-th group in the $l$-th transformer block, respectively. We set a constant hyper-parameter $\alpha$ as the coefficient of the balance loss. By encouraging a balanced group distribution, the system better leverages shared characteristics within each group while accounting for individual needs. It is worth noting that other methods exist to constrain the group gating network for even user allocation~\cite{shazeer2016outrageously,lepikhingshard}, with the balance loss presented here being one possible solution. Thus, the local foundation model loss for user $u$ is formulated as,
\begin{equation}\label{local_loss}
    \mathcal{L}_u = \mathcal{L}_{rec} + \alpha \cdot \mathcal{L}_{balance}
\end{equation}
The recommendation loss $\mathcal{L}{rec}$ is instantiated according to the task, such as binary cross-entropy for CTR prediction.


\begin{table*}[!t]
\setlength{\abovecaptionskip}{1mm}
\setlength{\belowcaptionskip}{-3mm}
\renewcommand{\arraystretch}{1.2}
\centering
\begin{tabular}{p{80pt}<{\centering}|p{65pt}||p{40pt}<{\centering}p{45pt}<{\centering}|p{40pt}<{\centering}p{45pt}<{\centering}|p{40pt}<{\centering}p{45pt}<{\centering}}
\hline 
\multirow{2}{*}{\textbf{Method}} & \textbf{Dataset} & \multicolumn{2}{c|}{\textbf{KuaiRand-Pure}} &
\multicolumn{2}{c|}{\textbf{KuaiSAR-R}} &
\multicolumn{2}{c}{\textbf{KuaiSAR-S}} \\
\cline{2-8}
& \textbf{Metric} & AUC $\uparrow$ & LogLoss $\downarrow$ & AUC $\uparrow$ & LogLoss $\downarrow$ & AUC $\uparrow$ & LogLoss $\downarrow$ \\
\hline
\multirow{2}{*}{\textbf{Non-Sequential}} & FedNCF & 0.7109 & 0.9793 & 0.6710 & 2.5670 & 0.5357 & 2.6338 \\
& FedPA & 0.6842 & 0.5935 & 0.6707 & 0.8055 & 0.5464 & 1.0655 \\
\hline
\multirow{9}{*}{\textbf{Transformer-based}} & FedSASRec & 0.7297 & 0.5181 & 0.7007 & 0.7105 & 0.5707 & 0.7034 \\
& \textbf{w/ \baby} & \textbf{0.7315*} & \textbf{0.4953*} & \textbf{0.7070*} & \textbf{0.6583*} & \textbf{0.5755*} & \textbf{0.5259*} \\
\cline{2-8}
& \multicolumn{1}{>{\columncolor{green!20}}l}{Improvement} & \multicolumn{1}{>{\columncolor{green!20}}c}{0.25\%} & \multicolumn{1}{>{\columncolor{green!20}}c}{4.60\%} & \multicolumn{1}{>{\columncolor{green!20}}c}{0.90\%} & \multicolumn{1}{>{\columncolor{green!20}}c}{7.93\%} & \multicolumn{1}{>{\columncolor{green!20}}c}{0.84\%} & \multicolumn{1}{>{\columncolor{green!20}}c}{33.75\%} \\
\cline{2-8}
& FedHSTU & 0.7304 & 0.5171 & 0.7030 & 0.7338 & 0.5725 & 0.7041 \\
& \textbf{w/ \baby} & \textbf{0.7330*} & \textbf{0.4949*} & \textbf{0.7064*} & \textbf{0.6426*} & \textbf{0.5735*} & \textbf{0.5987*} \\
\cline{2-8}
& \multicolumn{1}{>{\columncolor{green!20}}l}{Improvement} & \multicolumn{1}{>{\columncolor{green!20}}c}{0.36\%} & \multicolumn{1}{>{\columncolor{green!20}}c}{4.49\%} & \multicolumn{1}{>{\columncolor{green!20}}c}{0.48\%} & \multicolumn{1}{>{\columncolor{green!20}}c}{14.19\%} & \multicolumn{1}{>{\columncolor{green!20}}c}{0.17\%} & \multicolumn{1}{>{\columncolor{green!20}}c}{17.60\%} \\
\cline{2-8}
& FedLLaMA & 0.7302 & 0.5167 & 0.7003 & 0.7152 & 0.5724 & 0.7024 \\
& \textbf{w/ \baby} & \textbf{0.7311} & \textbf{0.4972*} & \textbf{0.7084*} & \textbf{0.6554*} & 0.\textbf{5787*} & \textbf{0.5730*} \\
\cline{2-8}
& \multicolumn{1}{>{\columncolor{green!20}}l}{Improvement} & \multicolumn{1}{>{\columncolor{green!20}}c}{0.12\%} & \multicolumn{1}{>{\columncolor{green!20}}c}{3.92\%} & \multicolumn{1}{>{\columncolor{green!20}}c}{1.16\%} & \multicolumn{1}{>{\columncolor{green!20}}c}{9.12\%} & \multicolumn{1}{>{\columncolor{green!20}}c}{1.10\%} & \multicolumn{1}{>{\columncolor{green!20}}c}{22.58\%} \\
\hline
\end{tabular}
\caption{Experimental results of baselines and our method on three datasets. ``\textbf{w/ \baby}" denotes enhancing the baseline with our proposed \baby \space and ``\textbf{Improvement}" indicates the performance gain achieved by integrating \baby. ``\textbf{{\Large *}}'' indicates statistically significant improvements (i.e., two-sided t-test with $p<0.05$) over the backbone baseline.}
\label{main_results}
\end{table*}

\subsection{Discussions about Practical Viability}
Unlike existing FFMs that fine-tune large-scale pre-trained models, \baby \space trains a lightweight foundation model from scratch on each client, effectively reducing deployment and computational overhead for resource-constrained devices. Furthermore, \baby \space keeps certain model parameters private, avoiding their upload to the server, thereby improving efficiency and enhancing system security. By reducing the number of parameters uploaded, communication overhead between the server and users is significantly reduced, which is particularly beneficial for recommender systems with large user bases. By keeping certain parameters local, the server has access to fewer publicly shared parameters, which reduces the risk of inferring users' private data from the model. To further enhance privacy protection, we explore integrating our approach with other privacy-preserving techniques, such as differential privacy~\cite{choi2018guaranteeing}, which will be discussed in detail in the experiments.

\section{Experiment}
\subsection{Experimental Setup}
We conduct experiments on three practical datasets: KuaiRand-Pure~\cite{gao2022kuairand}, KuaiSAR-R and KuaiSAR-S~\cite{Sun2023KuaiSAR}. This paper focuses on the click-through rate (CTR) prediction task and we adopt the prevalent \textit{\textbf{leave-one-out}} dataset split, following the setting in~\cite{kang2018self}. Model performance is assessed using two commonly used evaluation metrics: \textbf{\textit{AUC}} (Area Under the Curve) and \textbf{\textit{LogLoss}} (Logarithmic Loss). AUC evaluates ranking quality, and LogLoss assesses the calibration of predictions, offering a comprehensive evaluation of CTR prediction models.

\subsection{Baselines and Implementation Details}
\noindent \textbf{Baselines.}
\baby \space is an architecture-agnostic federated foundation recommendation framework, readily adaptable to common transformer-based recommendation models. To demonstrate its versatility, we select three representative backbone architectures: SASRec~\cite{kang2018self}, HSTU~\cite{zhaiactions}, and LLaMA~\cite{touvron2023llama}, and adapt them under the federated learning setting as the baselines, named \textbf{FedSASRec}, \textbf{FedHSTU} and \textbf{FedLLaMA}. SASRec is a well-established sequential recommendation model leveraging self-attention mechanisms. HSTU and LLaMA are state-of-the-art foundation model architectures. Additionally, we included two non-sequential recommendation frameworks, \textbf{FedNCF}~\cite{perifanis2022federated} and \textbf{FedPA}\cite{ijcai2024p603}, for further comparison. FedNCF is the first federated recommendation model based on a deep learning framework, while FedPA represents the leading federated recommendation method based on foundation models. 

\noindent \textbf{Implementation Details.}
For two non-sequential baselines FedNCF and FedPA, we fix batch size as 1,024. For three transformer-based baselines and our enhanced versions, we set transformer blocks as 2 for a fair comparison. Besides, we set the total communication rounds as 500 for all models to ensure convergence. All experiments are implemented using the PyTorch framework and repeated 5 times, with average results reported to ensure statistical reliability.

\begin{figure*}[!t]
\setlength{\abovecaptionskip}{0.1mm}
\setlength{\belowcaptionskip}{-2mm}
{\subfigure{\includegraphics[width=1.\linewidth]{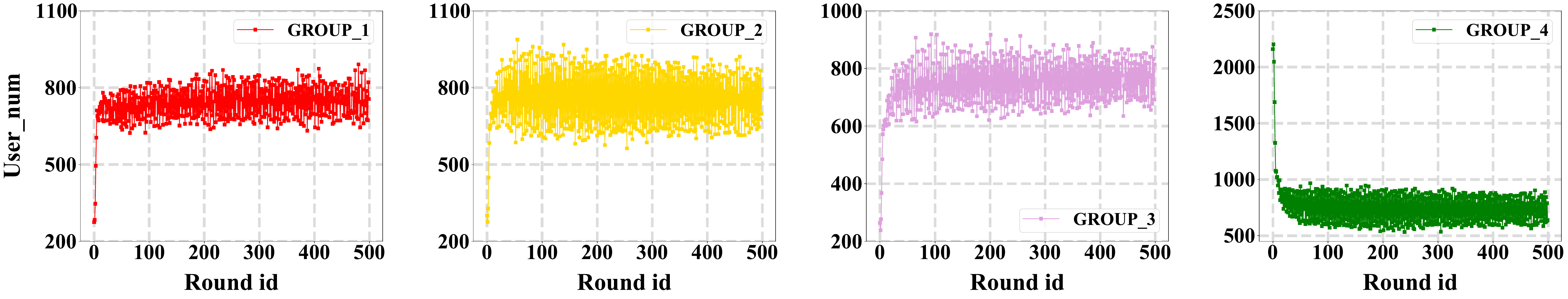}}}
{\subfigure{\includegraphics[width=1.\linewidth]{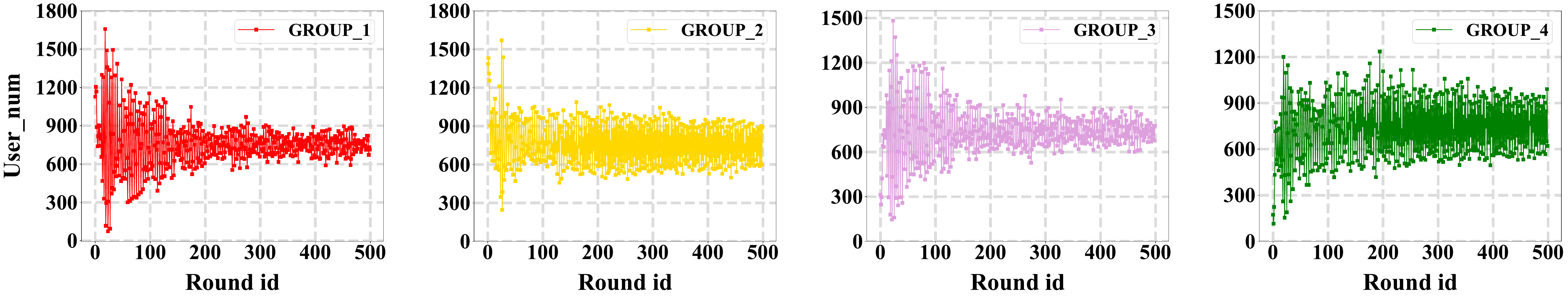}}}
\caption{Efficacy analysis of balance loss. The horizontal axis denotes the federated optimization rounds, and the vertical axis shows the number of users. The upper and lower subfigures display user grouping results for model’s two transformer blocks.}
\label{balance_loss}
\end{figure*}

\begin{figure*}[!t]
\setlength{\abovecaptionskip}{-1mm}
\setlength{\belowcaptionskip}{-3mm}
{\subfigure{\includegraphics[width=0.33\linewidth]{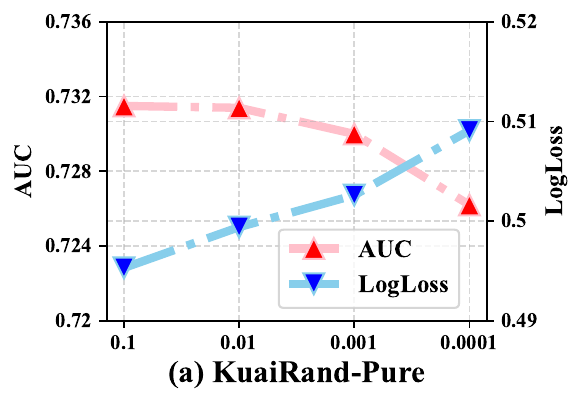}}}
{\subfigure{\includegraphics[width=0.33\linewidth]{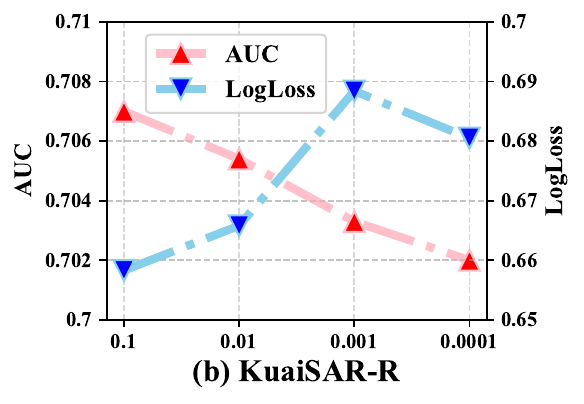}}}
{\subfigure{\includegraphics[width=0.33\linewidth]{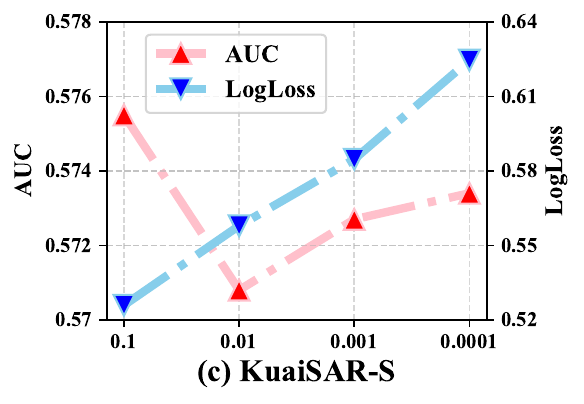}}}
    \caption{Impact of coefficient of the balance loss on model performace. }
    \label{alpha}
\end{figure*}
\subsection{Overall Performance}
Table~\ref{main_results} presents the evaluation results across three datasets using two performance metrics. We then provide a detailed discussion of the noteworthy observations and insights drawn from the experimental findings.

(1) \textbf{Transformer-based recommendation architectures outperform non-sequential models.} Compared to non-sequential models, which fail to fully capture the richness of user behavior, transformer-based models excel at identifying complex sequential patterns and user-item interactions. This underscores the importance of incorporating sequential information and modeling temporal dependencies to achieve high-quality recommendations. 

(2) \textbf{Our proposed method exhibits strong compatibility and versatility.} Integrating \baby \space into three distinct backbone architectures (FedSASRec, FedHSTU, and FedLLaMA) leads to significant performance improvements across the board. This showcases the model's ability to seamlessly integrate with and enhance various transformer-based recommendation frameworks.

(3) \textbf{Integrating our model with advanced foundation models typically results in improved performance.} Advanced foundation models, incorporating innovative advancements like rotational positional encoding (as seen in LLaMA), improve both efficiency and effectiveness. This suggests that combining our method with a broader range of advanced foundation models holds promise for achieving further performance gains.

\subsection{Efficacy Analysis of Balance Loss}
A key component in our \baby \space is the group gating network, which partitions users into groups to leverage their correlations for group-level personalization. However, skewed grouping can undermine effectiveness. To mitigate this, we develop a balance loss that encourages the group gating network to maintain uniform user routing. To verify the balance loss's effectiveness, we visualize user grouping dynamics during model training. Specifically, we track changes in user assignments across iterations, using the FedSASRec w/ \baby \space model on the KuaiRand-Pure dataset as an example.

As shown in Figure~\ref{balance_loss}, during model training, user allocation across groups gradually converges to a uniform distribution in both transformer blocks, with approximately 800 users assigned to each group. This uniform routing is crucial for ensuring that group-level personalization leverages user correlations equitably, without bias towards specific user populations. Experimental results show that the balance loss effectively encourages the group gating network to maintain an equitable user distribution.

\begin{figure*}[!t]
\setlength{\abovecaptionskip}{-2mm}
{\subfigure{\includegraphics[width=0.33\linewidth]{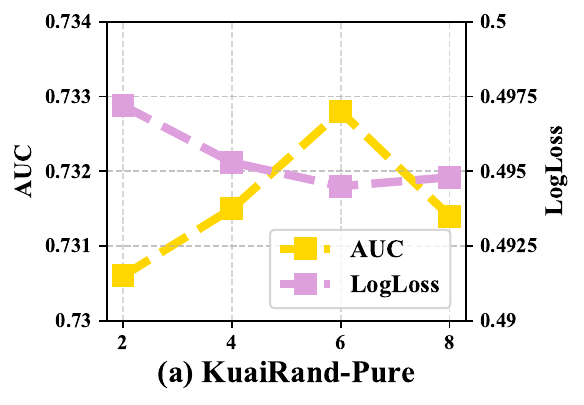}}}
{\subfigure{\includegraphics[width=0.33\linewidth]{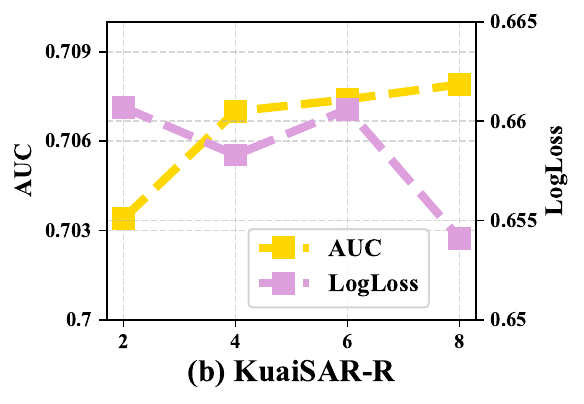}}}
{\subfigure{\includegraphics[width=0.33\linewidth]{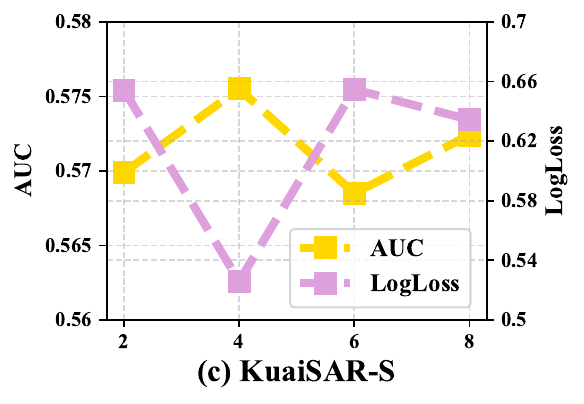}}}
    \caption{Impact of the number of user groups on model performance. }
    \label{beta}
\end{figure*}

\begin{table*}[!t]
\renewcommand\arraystretch{1.2}
\setlength{\abovecaptionskip}{1.5mm}
\setlength{\belowcaptionskip}{-2mm}
\centering
\small
\begin{tabular}{p{70pt}|p{40pt}<{\centering}||p{40pt}<{\centering}|p{40pt}<{\centering}|p{40pt}<{\centering}|p{40pt}<{\centering}|p{40pt}<{\centering}|p{40pt}<{\centering}}
\hline
\multirow{2}{*}{\textbf{Datasets}} & \multirow{2}{*}{\textbf{Metrics}} & \multicolumn{6}{c}{\textbf{Noise strength}} \\
\cline{3-8}
& & $0$ & \textbf{$0.1$} & \textbf{$0.2$} & \textbf{$0.3$} & \textbf{$0.4$} & \textbf{$0.5$}  \\
\hline
\multirow{2}{*}{KuaiRand-Pure}& AUC & \textbf{0.7315} & 0.7311 & 0.7307 & 0.7294 & 0.7293 & 0.7297 \\
& LogLoss & \textbf{0.4953} & 0.4960 & 0.4980 & 0.4978 & 0.4959 & 0.4993 \\
\hline
\multirow2{*}{KuaiSAR-R}& AUC & \textbf{0.7070} & 0.7039 & 0.7037 & 0.7037 & 0.7038 & 0.7037 \\
& LogLoss & 0.6583 & \textbf{0.6552} & 0.6733 & 0.6653 & 0.6622 & 0.6690 \\
\hline
\multirow2{*}{KuaiSAR-S}& AUC & \textbf{0.5755} & 0.5749 & 0.5708 & 0.5702 & 0.5718 & 0.5709 \\
& LogLoss & \textbf{0.5259} & 0.6042 & 0.6343 & 0.6457 & 0.6289 & 0.6271 \\
\hline
\end{tabular}
\caption{Results of privacy-preserving \baby \space with various noise strength.}
\label{ldp}
\end{table*}

\begin{table}[!t]
\renewcommand\arraystretch{1.2}
\setlength{\abovecaptionskip}{1.5mm}
\setlength{\belowcaptionskip}{-2mm}
\centering
\small
\begin{tabular}{p{52pt}|p{50pt}<{\centering}||p{30pt}<{\centering}|p{25pt}<{\centering}|p{25pt}<{\centering}}
\hline
\textbf{Datasets} & \textbf{Architectures} & \textbf{SASRec} & \textbf{HSTU} & \textbf{LLaMA} \\
\hline
\multirow{2}{*}{KuaiRand-Pure} & Parameter (k) & 59.78 & 67.89 & 63.19 \\
& Time (s) & 0.0733 & 0.0893 & 0.0890 \\
\hline
\multirow2{*}{KuaiSAR-R}& Parameter (k) & 52.72 & 62.53 & 57.83 \\
& Time (s) & 0.0614 & 0.0726 & 0.0503 \\
\hline
\multirow2{*}{KuaiSAR-S}& Parameter (k) & 51.75 & 62.74 & 58.03 \\
& Time (s) & 0.0616 & 0.0715 & 0.0698 \\
\hline
\end{tabular}
\caption{Summary of parameter size and testing time.}
\label{efficiency}
\end{table}

\subsection{Impact of Key Hyper-Parameters on Performance}
Our proposed \baby \space has two essential hyper-parameters, including the coefficient of the balancing loss and the number of user groups. We conduct evaluations on three datasets using model FedSASRec w/ \baby \space to empirically asses their impact on model performance. Specifically, we test the balance loss coefficient $\alpha$ with values in $\{0.0001, 0.001, 0.01, 0.1\}$, and the number of user groups $\beta$ with values in $\{2, 4, 6, 8\}$.

As shown in Figure~\ref{alpha}, the model achieves optimal performance across all three datasets when the balance loss coefficient is set to $\alpha = 0.1$. A larger coefficient is required to ensure that the group gating network maintains a and uniform user distribution, which in turn facilitates more effective group-level personalization.

In Figure~\ref{beta}, the model performance is observed to decline when the number of user groups $\beta$ is too small, such as $\beta = 2$. In this case, the limited granularity fails to capture the subtle differences in user behaviors. On the other hand, increasing the number of groups, such as setting $\beta = 6$, results in improved performance, as the finer user partitioning allows the model to better adapt to individual preferences. However, increasing the number of groups also increases the model's complexity and parameter size. We find that $\beta = 4$ strikes a good balance, providing enhanced performance without excessively inflating the model size. 

\subsection{Analysis about Practical Feasibility}
Deploying federated recommender systems in real-world presents significant challenges. User preferences are dynamic, requiring systems to provide timely and adaptive recommendations. The timeliness of these recommendations is crucial, as systems must deliver high-quality suggestions within a short time frame to meet practical needs. Additionally, real-world systems face a higher risk of malicious attacks, emphasizing the need for robust privacy protection. To evaluate the practical feasibility of our model, we evaluate it from two perspectives: efficiency and privacy.

To evaluate the efficiency of our method, we summarize the parameter size and testing time results across three transformer architectures on all datasets in Table~\ref{efficiency}. As highlighted in~\cite{ren2024advances}, existing FFMs typically integrate pre-trained foundation models with parameter scales in the hundreds of thousands. In contrast, each client in our approach only needs to deploy a model with approximately 60k parameters, with an average testing time of less than 0.1 seconds. These results demonstrate that our model offers significantly higher efficiency, making it better suited to meet the demands of real-world applications.

To enhance the privacy protection of our framework, we propose incorporating differential privacy techniques~\cite{choi2018guaranteeing}. Specifically, we introduce noise to the parameters uploaded from client devices to the server, reducing the risk of the server inferring private user information from parameter changes. As shown in Table~\ref{ldp}, our model maintains strong recommendation performance even with increased noise levels. This demonstrates that our approach not only ensures robust privacy protection but also satisfies the essential security requirements for practical deployment.

\section{Conclusion}
We present \baby, an innovative federated foundation model designed specifically for recommender systems. At its core, \baby \space trains a compact foundation model locally on each client device, avoiding the substantial computational overhead typically associated with large pre-trained models in federated settings. Additionally, we introduce a multifaceted user modeling mechanism that seamlessly integrates user-specific modeling while uncovering group-level similarities, enabling the development of an effective recommendation architecture that inherently preserves user privacy. Extensive experiments and in-depth analyses demonstrate significant performance improvements over advanced baselines, alongside superior compatibility. Moreover, comprehensive evaluations underscore the practical feasibility of our method, highlighting its efficiency and privacy-preserving capabilities.

\section{Acknowledgments}
Chunxu Zhang and Bo Yang are supported by the National Natural Science Foundation of China under Grant Nos. U22A2098, 62172185, 62206105 and 62202200; the Fundamental Research Funds for the Central Universities, JLU.

\bibliography{aaai25}

\begin{thebibliography}{47}
\providecommand{\natexlab}[1]{#1}

\bibitem[{Ba, Kiros, and Hinton(2016)}]{ba2016layer}
Ba, J.~L.; Kiros, J.~R.; and Hinton, G.~E. 2016.
\newblock Layer normalization.
\newblock \emph{arXiv preprint arXiv:1607.06450}.

\bibitem[{Bao et~al.(2023)Bao, Zhang, Wang, Zhang, Yang, Luo, Feng, He, and Tian}]{bao2023bi}
Bao, K.; Zhang, J.; Wang, W.; Zhang, Y.; Yang, Z.; Luo, Y.; Feng, F.; He, X.; and Tian, Q. 2023.
\newblock A bi-step grounding paradigm for large language models in recommendation systems.
\newblock \emph{arXiv preprint arXiv:2308.08434}.

\bibitem[{Charles et~al.(2024)Charles, Mitchell, Pillutla, Reneer, and Garrett}]{charles2024towards}
Charles, Z.; Mitchell, N.; Pillutla, K.; Reneer, M.; and Garrett, Z. 2024.
\newblock Towards federated foundation models: Scalable dataset pipelines for group-structured learning.
\newblock \emph{Advances in Neural Information Processing Systems}, 36.

\bibitem[{Chen et~al.(2023)Chen, Long, Shen, and Jiang}]{ijcai2023p393}
Chen, S.; Long, G.; Shen, T.; and Jiang, J. 2023.
\newblock Prompt Federated Learning for Weather Forecasting: Toward Foundation Models on Meteorological Data.
\newblock In \emph{Proceedings of the Thirty-Second International Joint Conference on Artificial Intelligence, {IJCAI-23}}, 3532--3540.

\bibitem[{Choi et~al.(2018)Choi, Tomei, Vicarte, Hanumolu, and Kumar}]{choi2018guaranteeing}
Choi, W.-S.; Tomei, M.; Vicarte, J. R.~S.; Hanumolu, P.~K.; and Kumar, R. 2018.
\newblock Guaranteeing local differential privacy on ultra-low-power systems.
\newblock In \emph{2018 ACM/IEEE 45th Annual International Symposium on Computer Architecture (ISCA)}, 561--574. IEEE.

\bibitem[{Du, Liu, and Wu(2021)}]{du2021modeling}
Du, Y.; Liu, H.; and Wu, Z. 2021.
\newblock Modeling multi-factor and multi-faceted preferences over sequential networks for next item recommendation.
\newblock In \emph{Machine Learning and Knowledge Discovery in Databases. Research Track: European Conference, ECML PKDD 2021, Bilbao, Spain, September 13--17, 2021, Proceedings, Part II 21}, 516--531. Springer.

\bibitem[{Gao et~al.(2022)Gao, Li, Zhang, Chen, Li, Lei, Jiang, and He}]{gao2022kuairand}
Gao, C.; Li, S.; Zhang, Y.; Chen, J.; Li, B.; Lei, W.; Jiang, P.; and He, X. 2022.
\newblock KuaiRand: An Unbiased Sequential Recommendation Dataset with Randomly Exposed Videos.
\newblock In \emph{Proceedings of the 31st ACM International Conference on Information and Knowledge Management}, CIKM '22, 3953–3957.

\bibitem[{Geng et~al.(2023)Geng, Tan, Liu, Fu, and Zhang}]{geng2023vip5}
Geng, S.; Tan, J.; Liu, S.; Fu, Z.; and Zhang, Y. 2023.
\newblock VIP5: Towards Multimodal Foundation Models for Recommendation.
\newblock In \emph{Findings of the Association for Computational Linguistics: EMNLP 2023}, 9606--9620.

\bibitem[{Hannon et~al.(2012)Hannon, McCarthy, O’Mahony, and Smyth}]{hannon2012multi}
Hannon, J.; McCarthy, K.; O’Mahony, M.~P.; and Smyth, B. 2012.
\newblock A multi-faceted user model for twitter.
\newblock In \emph{User Modeling, Adaptation, and Personalization: 20th International Conference, UMAP 2012, Montreal, Canada, July 16-20, 2012. Proceedings 20}, 303--309. Springer.

\bibitem[{Harte et~al.(2023)Harte, Zorgdrager, Louridas, Katsifodimos, Jannach, and Fragkoulis}]{harte2023leveraging}
Harte, J.; Zorgdrager, W.; Louridas, P.; Katsifodimos, A.; Jannach, D.; and Fragkoulis, M. 2023.
\newblock Leveraging large language models for sequential recommendation.
\newblock In \emph{Proceedings of the 17th ACM Conference on Recommender Systems}, 1096--1102.

\bibitem[{He et~al.(2016)He, Zhang, Ren, and Sun}]{he2016deep}
He, K.; Zhang, X.; Ren, S.; and Sun, J. 2016.
\newblock Deep residual learning for image recognition.
\newblock In \emph{Proceedings of the IEEE conference on computer vision and pattern recognition}, 770--778.

\bibitem[{Huang et~al.(2024)Huang, Yu, Xie, Zhang, Yao, and McAuley}]{huang2024foundation}
Huang, C.; Yu, T.; Xie, K.; Zhang, S.; Yao, L.; and McAuley, J. 2024.
\newblock Foundation models for recommender systems: A survey and new perspectives.
\newblock \emph{arXiv preprint arXiv:2402.11143}.

\bibitem[{Kang and McAuley(2018)}]{kang2018self}
Kang, W.-C.; and McAuley, J. 2018.
\newblock Self-attentive sequential recommendation.
\newblock In \emph{2018 IEEE international conference on data mining (ICDM)}, 197--206. IEEE.

\bibitem[{Lepikhin et~al.(2020)Lepikhin, Lee, Xu, Chen, Firat, Huang, Krikun, Shazeer, and Chen}]{lepikhingshard}
Lepikhin, D.; Lee, H.; Xu, Y.; Chen, D.; Firat, O.; Huang, Y.; Krikun, M.; Shazeer, N.; and Chen, Z. 2020.
\newblock GShard: Scaling Giant Models with Conditional Computation and Automatic Sharding.
\newblock In \emph{International Conference on Learning Representations}.

\bibitem[{Li(2021)}]{li2021leveraging}
Li, P. 2021.
\newblock Leveraging Multi-Faceted User Preferences for Improving Click-Through Rate Predictions.
\newblock In \emph{Proceedings of the 15th ACM Conference on Recommender Systems}, 864--868.

\bibitem[{Lin et~al.(2023)Lin, Dai, Xi, Liu, Chen, Zhang, Liu, Wu, Li, Zhu et~al.}]{lin2023can}
Lin, J.; Dai, X.; Xi, Y.; Liu, W.; Chen, B.; Zhang, H.; Liu, Y.; Wu, C.; Li, X.; Zhu, C.; et~al. 2023.
\newblock How can recommender systems benefit from large language models: A survey.
\newblock \emph{arXiv preprint arXiv:2306.05817}.

\bibitem[{Liu, Zhang, and Gulla(2023)}]{liu2023pre}
Liu, P.; Zhang, L.; and Gulla, J.~A. 2023.
\newblock Pre-train, Prompt, and Recommendation: A Comprehensive Survey of Language Modeling Paradigm Adaptations in Recommender Systems.
\newblock \emph{Transactions of the Association for Computational Linguistics}, 11: 1553--1571.

\bibitem[{Liu et~al.(2024)Liu, Zhao, Wang, Wang, Zhang, Sun, Li, Wang, Jia, Chen, Huang, and Tian}]{liu2024largelanguagemodelenhanced}
Liu, Q.; Zhao, X.; Wang, Y.; Wang, Y.; Zhang, Z.; Sun, Y.; Li, X.; Wang, M.; Jia, P.; Chen, C.; Huang, W.; and Tian, F. 2024.
\newblock Large Language Model Enhanced Recommender Systems: Taxonomy, Trend, Application and Future.
\newblock arXiv:2412.13432.

\bibitem[{Liu et~al.(2023)Liu, Guo, Liu, Tang, and Wang}]{liu2023user}
Liu, W.; Guo, W.; Liu, Y.; Tang, R.; and Wang, H. 2023.
\newblock User Behavior Modeling with Deep Learning for Recommendation: Recent Advances.
\newblock In \emph{Proceedings of the 17th ACM Conference on Recommender Systems}, 1286--1287.

\bibitem[{Long(2024)}]{ijcai2024p980}
Long, G. 2024.
\newblock The Rise of Federated Intelligence: From Federated Foundation Models Toward Collective Intelligence.
\newblock In \emph{Proceedings of the Thirty-Third International Joint Conference on Artificial Intelligence, {IJCAI-24}}, 8547--8552.

\bibitem[{McAuley and Leskovec(2013)}]{mcauley2013hidden}
McAuley, J.; and Leskovec, J. 2013.
\newblock Hidden factors and hidden topics: understanding rating dimensions with review text.
\newblock In \emph{Proceedings of the 7th ACM conference on Recommender systems}, 165--172.

\bibitem[{McMahan et~al.(2017)McMahan, Moore, Ramage, Hampson, and y~Arcas}]{mcmahan2017communication}
McMahan, B.; Moore, E.; Ramage, D.; Hampson, S.; and y~Arcas, B.~A. 2017.
\newblock Communication-efficient learning of deep networks from decentralized data.
\newblock In \emph{Artificial intelligence and statistics}, 1273--1282. PMLR.

\bibitem[{Miao et~al.(2025)Miao, Liu, Zhao, Zheng, Zhang, and Jensen}]{miao2025icde}
Miao, H.; Liu, Z.; Zhao, Y.; Zheng, K.; Zhang, Y.; and Jensen, C.~S. 2025.
\newblock LightTR: A Lightweight Framework for Federated Trajectory Recovery.
\newblock In \emph{ICDE}.

\bibitem[{Perifanis and Efraimidis(2022)}]{perifanis2022federated}
Perifanis, V.; and Efraimidis, P.~S. 2022.
\newblock Federated neural collaborative filtering.
\newblock \emph{Knowledge-Based Systems}, 242: 108441.

\bibitem[{Ren et~al.(2024)Ren, Yu, Peng, Tang, Li, Gao, Tan, Zhao, Li, Li et~al.}]{ren2024advances}
Ren, C.; Yu, H.; Peng, H.; Tang, X.; Li, A.; Gao, Y.; Tan, A.~Z.; Zhao, B.; Li, X.; Li, Z.; et~al. 2024.
\newblock Advances and open challenges in federated learning with foundation models.
\newblock \emph{arXiv preprint arXiv:2404.15381}.

\bibitem[{Shazeer et~al.(2016)Shazeer, Mirhoseini, Maziarz, Davis, Le, Hinton, and Dean}]{shazeer2016outrageously}
Shazeer, N.; Mirhoseini, A.; Maziarz, K.; Davis, A.; Le, Q.; Hinton, G.; and Dean, J. 2016.
\newblock Outrageously Large Neural Networks: The Sparsely-Gated Mixture-of-Experts Layer.
\newblock In \emph{International Conference on Learning Representations}.

\bibitem[{Srivastava et~al.(2014)Srivastava, Hinton, Krizhevsky, Sutskever, and Salakhutdinov}]{srivastava2014dropout}
Srivastava, N.; Hinton, G.; Krizhevsky, A.; Sutskever, I.; and Salakhutdinov, R. 2014.
\newblock Dropout: a simple way to prevent neural networks from overfitting.
\newblock \emph{The journal of machine learning research}, 15(1): 1929--1958.

\bibitem[{Sun et~al.(2023)Sun, Si, Zang, Leng, Niu, Song, Zhang, and Xu}]{Sun2023KuaiSAR}
Sun, Z.; Si, Z.; Zang, X.; Leng, D.; Niu, Y.; Song, Y.; Zhang, X.; and Xu, J. 2023.
\newblock KuaiSAR: A Unified Search And Recommendation Dataset.
\newblock In \emph{Proceedings of the 32nd ACM International Conference on Information and Knowledge Management}.

\bibitem[{Touvron et~al.(2023)Touvron, Martin, Stone, Albert, Almahairi, Babaei, Bashlykov, Batra, Bhargava, Bhosale et~al.}]{touvron2023llama}
Touvron, H.; Martin, L.; Stone, K.; Albert, P.; Almahairi, A.; Babaei, Y.; Bashlykov, N.; Batra, S.; Bhargava, P.; Bhosale, S.; et~al. 2023.
\newblock Llama 2: Open foundation and fine-tuned chat models.
\newblock \emph{arXiv preprint arXiv:2307.09288}.

\bibitem[{Wu et~al.(2023)Wu, Zheng, Qiu, Wang, Gu, Shen, Qin, Zhu, Zhu, Liu et~al.}]{wu2023survey}
Wu, L.; Zheng, Z.; Qiu, Z.; Wang, H.; Gu, H.; Shen, T.; Qin, C.; Zhu, C.; Zhu, H.; Liu, Q.; et~al. 2023.
\newblock A survey on large language models for recommendation.
\newblock \emph{arXiv preprint arXiv:2305.19860}.

\bibitem[{Wu et~al.(2024)Wu, Zheng, Qiu, Wang, Gu, Shen, Qin, Zhu, Zhu, Liu et~al.}]{wu2024survey}
Wu, L.; Zheng, Z.; Qiu, Z.; Wang, H.; Gu, H.; Shen, T.; Qin, C.; Zhu, C.; Zhu, H.; Liu, Q.; et~al. 2024.
\newblock A survey on large language models for recommendation.
\newblock \emph{World Wide Web}, 27(5): 60.

\bibitem[{Xu et~al.(2024)Xu, Miao, Wang, Yu, and Wang}]{xu2024pefad}
Xu, R.; Miao, H.; Wang, S.; Yu, P.~S.; and Wang, J. 2024.
\newblock PeFAD: a parameter-efficient federated framework for time series anomaly detection.
\newblock In \emph{SIGKDD}, 3621--3632.

\bibitem[{Yan and Long(2023)}]{yan2023personalization}
Yan, P.; and Long, G. 2023.
\newblock Personalization disentanglement for federated learning.
\newblock In \emph{2023 IEEE International Conference on Multimedia and Expo (ICME)}, 318--323. IEEE.

\bibitem[{Yu, Mu{\~n}oz, and Jannesari(2023)}]{yu2023federated}
Yu, S.; Mu{\~n}oz, J.~P.; and Jannesari, A. 2023.
\newblock Federated foundation models: Privacy-preserving and collaborative learning for large models.
\newblock \emph{arXiv preprint arXiv:2305.11414}.

\bibitem[{Yu, Munoz, and Jannesari(2024)}]{yu2024federated}
Yu, S.; Munoz, J.~P.; and Jannesari, A. 2024.
\newblock Federated Foundation Models: Privacy-Preserving and Collaborative Learning for Large Models.
\newblock In \emph{Proceedings of the 2024 Joint International Conference on Computational Linguistics, Language Resources and Evaluation (LREC-COLING 2024)}, 7174--7184.

\bibitem[{Zhai et~al.(2024{\natexlab{a}})Zhai, Liao, Liu, Wang, Li, Cao, Gao, Gong, Gu, He et~al.}]{zhaiactions}
Zhai, J.; Liao, L.; Liu, X.; Wang, Y.; Li, R.; Cao, X.; Gao, L.; Gong, Z.; Gu, F.; He, J.; et~al. 2024{\natexlab{a}}.
\newblock Actions Speak Louder than Words: Trillion-Parameter Sequential Transducers for Generative Recommendations.
\newblock In \emph{Forty-first International Conference on Machine Learning}.

\bibitem[{Zhai et~al.(2024{\natexlab{b}})Zhai, Liao, Liu, Wang, Li, Cao, Gao, Gong, Gu, He et~al.}]{zhai2024actions}
Zhai, J.; Liao, L.; Liu, X.; Wang, Y.; Li, R.; Cao, X.; Gao, L.; Gong, Z.; Gu, F.; He, M.; et~al. 2024{\natexlab{b}}.
\newblock Actions speak louder than words: Trillion-parameter sequential transducers for generative recommendations.
\newblock \emph{arXiv preprint arXiv:2402.17152}.

\bibitem[{Zhang et~al.(2024{\natexlab{a}})Zhang, Chen, Sheng, Wang, and Chua}]{zhang2024generative}
Zhang, A.; Chen, Y.; Sheng, L.; Wang, X.; and Chua, T.-S. 2024{\natexlab{a}}.
\newblock On generative agents in recommendation.
\newblock In \emph{Proceedings of the 47th International ACM SIGIR Conference on Research and Development in Information Retrieval}, 1807--1817.

\bibitem[{Zhang et~al.(2024{\natexlab{b}})Zhang, Long, Guo, Fang, Song, Liu, Zhou, Zhang, Liu, and Yang}]{ijcai2024p603}
Zhang, C.; Long, G.; Guo, H.; Fang, X.; Song, Y.; Liu, Z.; Zhou, G.; Zhang, Z.; Liu, Y.; and Yang, B. 2024{\natexlab{b}}.
\newblock Federated Adaptation for Foundation Model-based Recommendations.
\newblock In \emph{Proceedings of the Thirty-Third International Joint Conference on Artificial Intelligence, {IJCAI-24}}, 5453--5461.

\bibitem[{Zhang et~al.(2024{\natexlab{c}})Zhang, Hou, Xie, Sun, McAuley, Zhao, Lin, and Wen}]{zhang2024agentcf}
Zhang, J.; Hou, Y.; Xie, R.; Sun, W.; McAuley, J.; Zhao, W.~X.; Lin, L.; and Wen, J.-R. 2024{\natexlab{c}}.
\newblock Agentcf: Collaborative learning with autonomous language agents for recommender systems.
\newblock In \emph{Proceedings of the ACM on Web Conference 2024}, 3679--3689.

\bibitem[{Zhang et~al.(2023)Zhang, Xie, Hou, Zhao, Lin, and Wen}]{zhang2023recommendation}
Zhang, J.; Xie, R.; Hou, Y.; Zhao, W.~X.; Lin, L.; and Wen, J.-R. 2023.
\newblock Recommendation as instruction following: A large language model empowered recommendation approach.
\newblock \emph{arXiv preprint arXiv:2305.07001}.

\bibitem[{Zhao et~al.(2024)Zhao, Fan, Li, Liu, Mei, Wang, Wen, Wang, Zhao, Tang et~al.}]{zhao2024recommender}
Zhao, Z.; Fan, W.; Li, J.; Liu, Y.; Mei, X.; Wang, Y.; Wen, Z.; Wang, F.; Zhao, X.; Tang, J.; et~al. 2024.
\newblock Recommender systems in the era of large language models (llms).
\newblock \emph{IEEE Transactions on Knowledge and Data Engineering}.

\bibitem[{Zheng et~al.(2024{\natexlab{a}})Zheng, Chao, Qiu, Zhu, and Xiong}]{zheng2024harnessing}
Zheng, Z.; Chao, W.; Qiu, Z.; Zhu, H.; and Xiong, H. 2024{\natexlab{a}}.
\newblock Harnessing large language models for text-rich sequential recommendation.
\newblock In \emph{Proceedings of the ACM on Web Conference 2024}, 3207--3216.

\bibitem[{Zheng et~al.(2024{\natexlab{b}})Zheng, Hu, Gao, Zhu, and Xiong}]{zheng2024mirror}
Zheng, Z.; Hu, X.; Gao, S.; Zhu, H.; and Xiong, H. 2024{\natexlab{b}}.
\newblock Mirror: A multi-view reciprocal recommender system for online recruitment.
\newblock In \emph{Proceedings of the 47th International ACM SIGIR Conference on Research and Development in Information Retrieval}, 543--552.

\bibitem[{Zheng et~al.(2024{\natexlab{c}})Zheng, Hu, Qiu, Cheng, Gao, Song, Zhu, and Xiong}]{zheng2024bilateral}
Zheng, Z.; Hu, X.; Qiu, Z.; Cheng, Y.; Gao, S.; Song, Y.; Zhu, H.; and Xiong, H. 2024{\natexlab{c}}.
\newblock Bilateral Multi-Behavior Modeling for Reciprocal Recommendation in Online Recruitment.
\newblock \emph{IEEE Transactions on Knowledge and Data Engineering}.

\bibitem[{Zhou et~al.(2023)Zhou, Cao, Huang, Ye, Zhang, Liu, Xie, Hua, and Kim}]{zhou2023exploring}
Zhou, P.; Cao, M.; Huang, Y.-L.; Ye, Q.; Zhang, P.; Liu, J.; Xie, Y.; Hua, Y.; and Kim, J. 2023.
\newblock Exploring recommendation capabilities of gpt-4v (ision): A preliminary case study.
\newblock \emph{arXiv preprint arXiv:2311.04199}.

\bibitem[{Zhuang, Chen, and Lyu(2023)}]{zhuang2023foundation}
Zhuang, W.; Chen, C.; and Lyu, L. 2023.
\newblock When foundation model meets federated learning: Motivations, challenges, and future directions.
\newblock \emph{arXiv preprint arXiv:2306.15546}.

\end{thebibliography}

\end{document}